\documentclass{PoS}

\usepackage{graphicx}
\usepackage{amssymb,amsmath}

\newcommand{\bea}{\begin{eqnarray}}
\newcommand{\eea}{\end{eqnarray}}

\newcommand{\del}{\partial}
\newcommand{\Nred}{N_{\rm red}}
\title{A filtering technique for the temporally reduced matrix of the Wilson fermion determinant}
\ShortTitle{Filtering technique for the temporally reduced matrix}

\author{Yasunori Futamura\\
   Department of computer science, Tsukuba University\\
        E-mail: \email{futamura@mma.cs.tsukuba.ac.jp}
}

\author{Shoji Hashimoto\\
KEK, Tsukuba 305-0801, Japan \\
        E-mail: \email{shoji.hashimoto@kek.jp}
}

\author{Akira Imakura\\
   Department of computer science, Tsukuba University\\
         E-mail: \email{imakura@cs.tsukuba.ac.jp}
}

\author{\speaker{Keitaro Nagata}\thanks{KEK-CP-314}\\
KEK, Tsukuba 305-0801, Japan\\
        E-mail: \email{knagata@post.kek.jp}
}

\author{Tetsuya Sakurai\\
   Department of computer science, Tsukuba University\\
     E-mail: \email{sakurai@cs.tsukuba.ac.jp}
}

\abstract{
The Wilson fermion determinant can be written in the form of a series
expansion in fugacity $\xi=\exp(\mu/T)$, provided that the eigenmodes of the
temporally reduced operator are obtained.  Since the calculation of
all eigenmodes rapidly becomes prohibitive for larger volumes, we
develop a method to calculate only the low-energy eigenmodes of the
reduced matrix using a matrix filetering technique. 
This provides a basis for an approximation to neglect uninteresting ultraviolet contributions.
}

\FullConference{The 32nd International Symposium on Lattice Field Theory,\\
		23-28 June, 2014\\
		Columbia University New York, NY}

\begin{document}

\section{Introduction}
Simulation of finite quark density system is a long-standing challenge in lattice QCD simulations. 
Since the ordinary Monte Carlo technique breaks down, several approaches are being 
pursued to overcome the sign problem. 
One example is a temporal reduction formula of the fermion determinant, 
which reexpresses the fermion determinant as a series expansion in terms of 
fugacity~\cite{Gibbs:1986hi,Hasenfratz:1991ax,Adams:2003rm,Borici:2004bq,Nagata:2010xi,Alexandru:2010yb}. 
The formula involves with a temporally reduced matrix, which we simply refer to as a reduced matrix. 
The fermion determinant is reduced to a form analytic with regard to quark chemical potential, 
provided the eigenvalues of the reduced matrix are obtained. 
Thus, the formula offers an useful tool to study chemical potential dependence of the fermion determinant~\cite{Barbour:1988ax,Barbour:1991vs,Hasenfratz:1991ax,deForcrand:2006ec,Kratochvila:2005mk,Kratochvila:2004wz,Li:2010qf,Nagata:2012tc,Nagata:2014bra}. 

In spite of its theoretical advantages,  the use of the reduction formula has so far 
been limited to small lattice volumes due to the numerical cost to calculate 
the eigenvalues of the reduced matrix, which is a dense complex non-symmetric matrix. 
Through the previous studies~\cite{Gibbs:1986hi,Fodor:2007ga,Nagata:2012tc,Nagata:2012mn}, 
the spectral properties of the reduced matrix has been partly revealed. 
Although the eigenvalues of the reduced matrix span a wide range of magnitude, 
physically relevant eigen-modes are located near the unit circle on the complex plane, 
which are of medium magnitude. 

In this work, we propose a method to treat only the physical eigen-modes. 
For this purpose, we employ an algorithm proposed by one of the present author (TS) and Sugiura. 
The algorithm is based on a contour integral, and enables to obtain only desired eigenvalues 
by adjusting integral contours. 
We report first exploratory study of the algorithm applied to the temporally reduced matrix. 
This paper is organized as follows. 
The reduced matrix is introduced in the next section.
The algorithm of eigenvalue calculation is explained in section~\ref{sec:SSmethod}. 
Numerical results are given in section~\ref{sec:Result}.
Final section is devoted to a summary. 
\section{Reduced matrix}
\label{sec:ReducedMatrix}

We consider the Wilson fermion matrix defined by 
\begin{align}
\Delta(x,x') &= \delta_{x,x'}
 - \kappa \sum_{i=1}^{3} \left\{
        (r-\gamma_i) U_i(x) \delta_{x',x+\hat{i}}
      + (r+\gamma_i) U_i^{\dagger}(x') \delta_{x',x-\hat{i}} \right\}
\nonumber \\
  & - \kappa \left\{
        e^{+\mu a}(r-\gamma_4) U_4(x) \delta_{x',x+\hat{4}}
      + e^{-\mu a}(r+\gamma_4) U_4^{\dagger}(x') \delta_{x',x-\hat{4}}
\right\} - \delta_{x, x^\prime} C_{SW} \kappa \sum_{\mu \le \nu} \sigma_{\mu\nu} 
F_{\mu\nu}.
\label{Wfermion}
\end{align}
where $r$ and $C_{SW}$ are coefficients of the Wilson term and the clover term. 
$a$, $\kappa$ and $\mu$ are lattice spacing, hopping parameter and chemical potential, respectively.
We divide Eq.~(\ref{Wfermion}) into three terms, according to its temporal structure, as 
\begin{align}
\Delta =  B -\kappa \left[ e^{+\mu a} (1-\gamma_4) U_4(x) \delta_{x^\prime, x+\hat{4}} 
+e^{-\mu a} (1+\gamma_4) U^\dagger_4(x^\prime) \delta_{x^\prime, x-\hat{4}}\right].
\end{align}
$B$ is the spatial part of the Wilson fermion matrix,  
\begin{align}
 B  =  \delta_{x, x^\prime} -\kappa \sum_{i=1}^{3} \Bigl[
(1-\gamma_i) U_i(x) \delta_{x^\prime, x+\hat{i}}  
  + (1+\gamma_i) U_i^\dagger(x^\prime) \delta_{x^\prime, x-\hat{i}}\Bigr] 
- \kappa  C_{SW} \delta_{x, x^\prime}  \sum_{\mu \le \nu} \sigma_{\mu\nu} 
F_{\mu\nu}.
\label{Jul202011eq1}
\end{align}
We introduce two block-matrices
\begin{subequations}
\begin{align}
(\alpha_i)^{ab} =& B^{ab}(\vec{x}, \vec{y}, t_i) \; r_{-}
         -2  \kappa \; r_{+} \delta^{ab} \delta_{\vec{x},\vec{y}}, \\
(\beta_i)^{ab} =& \Bigl[ B^{ac}(\vec{x}, \vec{y}, t_i)\; r_{+}
-2 \kappa \; r_{-} \delta_{\vec{x},\vec{y}} \delta^{ac}\Bigr] U_4^{cb}(\vec{y}, t_i).
\label{Eq:2012Feb21eq1}
\end{align}%
\end{subequations}%
where $a, b$, and $c$ are color indices. 
$r_\pm = (1 \pm \gamma_4)/2$ are projection operators. 
$\alpha_i$ describes a spatial hop of a quark at $t=t_i$, while 
$\beta_i$ describes a spatial hop at $t=t_i$ as well as a temporal hop 
to the next time slice. 
The reduced matrix is defined by 
\begin{align}
Q = (\alpha_1^{-1} \beta_1) \cdots (\alpha_{N_t}^{-1} \beta_{N_t}), 
\end{align}
where ${\rm rank}(Q)=N_{\rm red}=4 N_c N_x N_y N_z$. 
It is smaller than ${\rm rank}(\Delta)= 4 N_c N_x N_y N_z N_t$.
Here $N_x, N_y,$, $N_z$ and $N_t$ are the numbers of lattice sites in each direction, 
and $N_c$ is the number of colors. 
Using the reduced matrix, the fermion determinant is rewritten as
\begin{align}
\det \Delta ( \mu) = C_0 \det (Q + \xi), 
\end{align}
where $\xi=\exp(-\mu/T)$. 
Using the eigenvalues of $Q$, $\det \Delta(\mu)$ can be rewritten as 
\begin{align}
\det \Delta(\mu) = C_0 \prod_{n=1}^{N_{\rm red}} (\lambda_n + \xi).
\label{Eq:2014Oct29eq2}
\end{align}
Expanding the product, Eq.~(\ref{Eq:2014Oct29eq2}) reads as 
a series expansion in terms of the fugacity $\xi$. 
This formula gives $\det \Delta(\mu)$ for any values 
of $\mu$ on a given background gauge configuration.


The reduced matrix $Q$ has $N_{\rm red}$ eigenvalues. 
They are complex, and appear as pairs $(\lambda_n, 1/\lambda_n^*)$. 
There are evidences that eigenvalues near the unit circle ($|\lambda_n|\sim 1$) correspond 
to the physical modes contributing to low energy physics. 
For instance, the masses of pions and other ground state hadrons are dominated by 
eigenvalues near the unit circle~\cite{Gibbs:1986hi,Fodor:2007ga}.
The reduced matrix $Q$ describes a temporal quark line and interpreted as 
a generalization of the Polyakov line. 
Similar to the Polyakov line, the eigenvalues are parameterized as 
$\lambda_n = \exp(-\epsilon_n / T - i\theta_n)$~\cite{Nagata:2012tc}. 
Using this scaling behavior, the number operator is written as 
\begin{align}
\hat{n}&=\frac{T}{V_s} (\det \Delta(\mu))^{-1} \frac{\del\det \Delta(\mu)}{\del \mu}
 = \frac{1}{V_s}\sum_{n=1}^{\Nred/2}
\left( 
\frac{\lambda_n \xi^{-1}}{1+\lambda_n\xi^{-1}} 
-\frac{\lambda_n^*\xi}{1+\lambda_n^*\xi}
\right) \nonumber \\
&= \frac{1}{V_s} \sum_{n=1}^{\Nred/2}
\left(
\frac{1}{1+e^{(\epsilon_n-\mu)/T - i\theta_n}}
-\frac{1}{1+e^{(\epsilon_n+\mu)/T + i\theta_n}}
\right).
\label{Eq:2013Sep28eq1}
\end{align}
This is the same form as a Fermi distribution on a given background gauge configuration. 
$\epsilon_n$ is interpreted as a single energy level of a quark for a 
given background gauge configuration. 
The quark number density is dominated by small $\epsilon$ corresponding to 
$|\lambda|\sim 1$.
It follows from these understandings that among all the eigenvalues, 
physical eigen-modes contributing to low energy physics are located near 
the unit circle on the complex plane. 
They are of medium magnitude, while small and large eigenvalues are 
high energy eigen-modes related to ultraviolet physics. 

\section{Algorithm to calculate relevant eigenvalues}
\label{sec:SSmethod}
Numerical difficulty of the eigenvalue calculation hinders the application of 
the reduction formula to large lattice volume.
To circumvent the problem, we propose to treat only low energy eigenvalues.
As we have discussed in the previous section, low energy eigen-modes are located near the 
unit circle and they are of medium magnitude. 
Thus, we need an eigensolver to extract medium eigenvalues.
In this work, we employ the algorithm proposed in ~\cite{Sakurai:2002aaa,Sakurai:2012aaa}. 
It is based on a contour integral, and enables to obtain desired eigenvalues 
included inside a given contour. 
There are single and blocked versions of the algorithm, and in this work we apply the 
blocked version, which we refer to as blocked Sakurai-Sugiura (bSS) method. 

We consider an eigenvalue problem
\begin{align}
A x_n = \lambda_n x_n, (n=1, 2, \cdots n)
\label{Eq:2014Oct29eq1}
\end{align}
where $A \in \mathbb{C}^{n\times n}$ is a complex matrix of rank $n$. 
$\lambda_n$, and $x_n$ are eigenvalues and eigenvectors.

Let us consider $m$-eigenvalues located inside a contour $\Gamma$. 
First, we define matrices $S_0$, $S_1$, $\cdots$ as 
\begin{align}
S_k &= \frac{1}{2\pi i } \int_\Gamma z^k (z - A)^{-1} V d z, \;\; (k = 0, 1, \cdots, M-1  ), 
\label{Eq:2014Oct30eq1}
\end{align}
where $\Gamma$ is a closed path on a complex $z$ plane. $V$ includes $L$ random-vectors
\begin{align}
V & = \{ v_1, v_2, \cdots, v_L\} \in \mathbb{R}^{n \times L}.
\end{align}
$M$ and $L$ are integers chosen such that $L M \ge m$.
Next, we perform the singular-value decomposition for a rectangular matrix 
$S=[S_0, S_1, \cdots, S_{M-1}]\in \mathbb{C}^{n \times ML}$ as
\begin{subequations}
\begin{align}
S&= U \Sigma W^\dagger, \\
\Sigma &= {\rm diag} (\sigma_1, \sigma_2, \cdots, \sigma_{LM}), \\
U & = (u_1, u_2, \cdots, u_{LM}) \in \mathbb{C}^{n \times LM}.
\end{align}
\end{subequations}

Now, we project the original eigen problem to a smaller one by using first $l$-vectors in $U$. 
First, we define $U_l = (u_1, u_2, \cdots u_l)$, and calculate $A_l = U_l^\dagger A U_l, $
where ${\rm rank}(A_l) = l$. 
Here, the value of $l$ is determined so that singular values satisfy
$\sigma_i> \epsilon_{\rm SVD}, \; (i \le l)$ for a given cut $\epsilon_{\rm SVD}$.
We denote the eigenvalues and eigenvectors for $A_l$ as $\omega_i$, and $r_i, \; (i=1, 2, \cdots l)$ : 
\begin{align}
A_l r_j & = \omega_j r_j. 
\end{align}
They are related to eigenpairs of the original eigen problem 
\begin{subequations}
\begin{align}
\lambda_j &= \omega_j, \\
x_j &= U_m r_j.
\end{align}
\end{subequations}

The bSS method involves the matrix inversion $(z-A)^{-1}$ in Eq.~(\ref{Eq:2014Oct30eq1}). 
The most efficient way to achieve this step is to use an iterative
solver with a shifted algorithm, such as the shifted BiCGStab. 
Although this is an ideal case, we, however, encounter an ill-conditioned problem 
in the application of the BiCGStab algorithm to the reduced matrix $Q$ because of 
a dense property of $Q$.
We are forced to use a direct method; we construct the matrix $Q$ 
and calculate $(z-Q)^{-1}$ in a direct method. 
For more efficient calculation, we need an iterative solver for $(z-Q)^{-1}$. 
We leave this for future studies. 


\section{Simulation and Result}
\label{sec:Result}
Now, we apply the bSS method to the reduced matrix. 
For this exploratory study, we take the following setup: 
a configuration for the clover-improved Wilson fermion and renormalization-group 
improved gauge action on the lattice volume $(N_x, N_y, N_z, N_t)=(4,4,4,4)$.
To extract low-lying modes near the unit circle, we divide the domain surrounded by 
the unit circle into some rings, each of which is surrounded by two circles. 
Denoting two contours as $\Gamma_1$ and $\Gamma_2$, the eigenvalues inside them 
are obtained by replacing $\int_\Gamma$ with $\int_{\Gamma_1} - \int_{\Gamma_2}$ in Eq.~(\ref{Eq:2014Oct30eq1}).
To make a comparison, we also calculate eigenvalues using ZGEES subroutine in LAPACK . 
We also need to determine parameters used in the bSS method:
the number of vectors $L$, the number of integral points $N_{\rm int}$, the maximum order of moments $M$
and $l$, which is the number of vectors $u_i$ taken into account for the projection to the small 
eigen problem. 

\begin{figure}[htbp] 
\includegraphics[width=9cm]{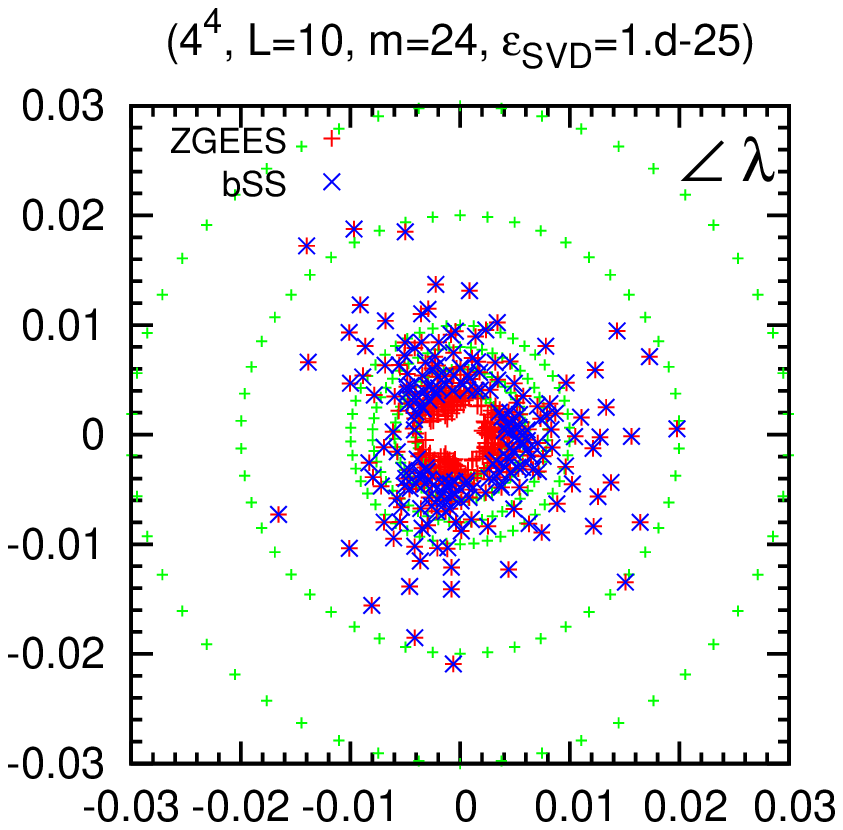}%
\includegraphics[width=8cm]{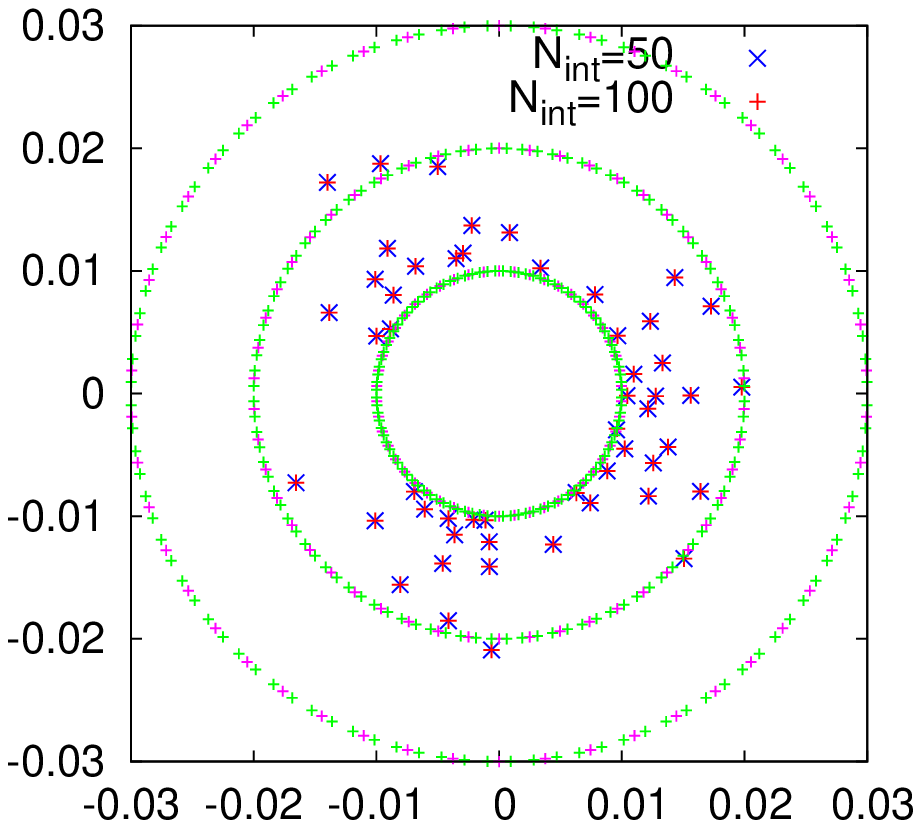}%
\caption{Left : Eigenvalues obtained from blocked SS method (blue cross) 
compared with those obtained from ZGEES subroutine in LAPACK (red plus). 
Green symbols denote integral points, and eigenvalues are obtained for 
each domain. Right panel : results for two different integral points. }
\label{Fig:2014Oct29fig3}
\end{figure} 
The left panel of Fig.~\ref{Fig:2014Oct29fig3} shows the eigenvalue distribution, 
where eigenvalues are obtained for a corresponding ring domain by which they are surrounded. 
The right panel shows the comparison of results obtained for two different values of integral points $N_{\rm int}$.
As they precisely agree, the integral points $N_{\rm int}=50$ are sufficient for the present case. 
The method works well for this case, in spite of the fact that some eigenvalues
are located close to the integral contours.

\begin{figure}[htbp] 
\includegraphics[width=8cm]{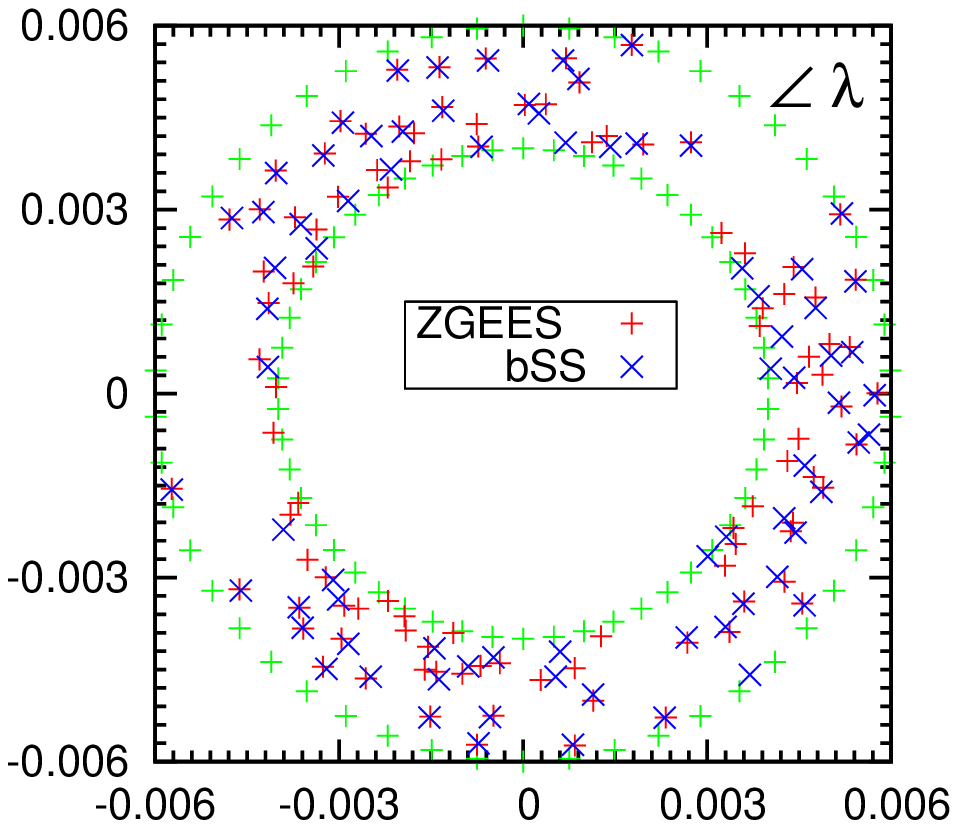}%
\includegraphics[width=8cm]{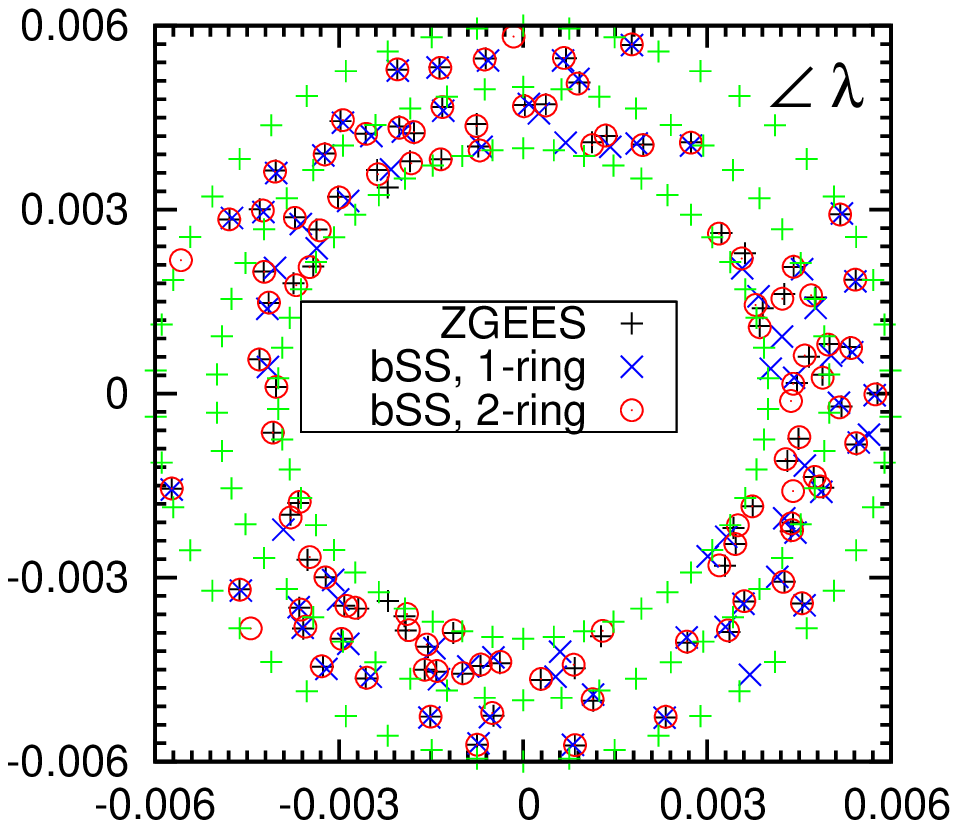}
\includegraphics[width=8cm]{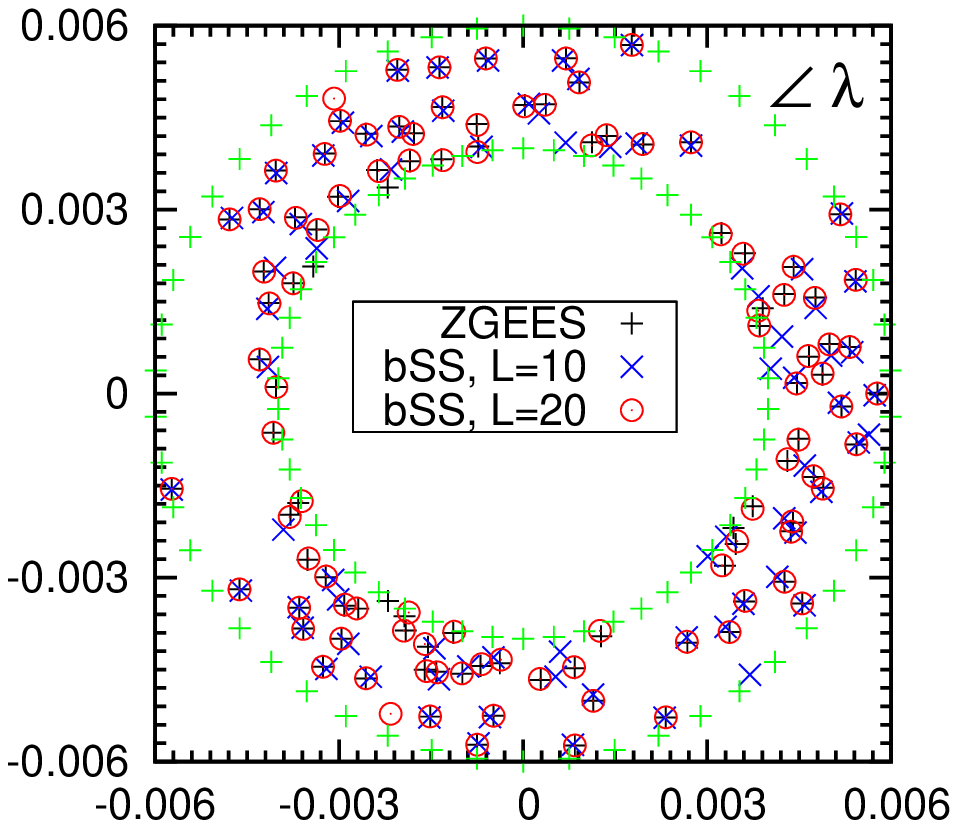}%
\includegraphics[width=8cm]{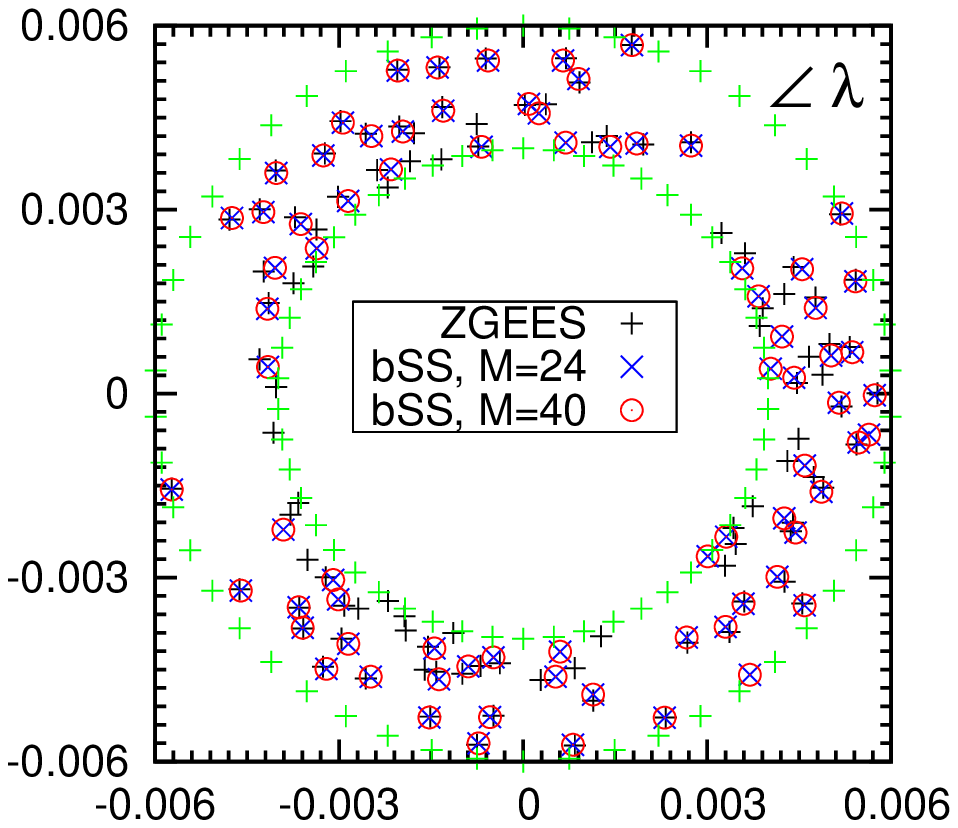}%
\caption{Eigenvalue distribution in the innermost ring domain for several cases. 
Top left : $(L,M)=(10,24)$. Top right : Division into two rings with $(L,M)=(10,24)$. 
Bottom left : $L$-dependence. Bottom right : $M$-dependence.}
\label{Fig:2014Oct29fig1}
\end{figure} 
Figure~\ref{Fig:2014Oct29fig1} shows the eigenvalue distribution in the innermost ring.
The top left panel of Fig.~\ref{Fig:2014Oct29fig1} is the magnification of 
Fig.~\ref{Fig:2014Oct29fig3}. 
We found that some eigenvalues are not obtained correctly in the innermost ring. 
The precision of the method is sensitive to the density of eigenvalues 
in a given domain.
There may be some possibilities to improve the precision for the dense domain. 
We have examined thinner division of the domain (top right), 
increase the number of vectors $L$ (bottom left), order of moments $M$ (bottom right). 
We found that the increase of the maximum moment $M$ does not improve the result so much. 
On the other hand, some eigenvalues are correctly reproduced 
with the increase of the number of vectors or thinner division. 
However,  some eigenvalues are not reproduced correctly even with 
the two improved cases. Such ill case seems to appear near the contours.
Other possibility is to increase the number of integral points
or to decrease the criterion $\epsilon_{\rm SVD}$. 
As we have shown, the result is not sensitive to the number of integral points. 
We confirmed that $\epsilon_{\rm SVD}=10^{-25}$ is sufficiently small for the 
present case and the result does not change by the decrease of $\epsilon_{\rm SVD}$.

\section{Summary}
We have considered the extraction of the physical modes of the temporally 
reduced Wilson fermion matrix by using the bSS method, which is 
an eigensolver based on the contour integral.
We numerically test the application of the bSS method for 
the reduced matrix. The method reproduces the eigenvalues for 
sparse domains, while it fails to reproduce some eigenvalues in dense domains.
The precision of the method, thus, depends on the density of the eigenvalues 
for a given domain. 
Such cases would be solved correctly by adjusting some parameters and contours 
in the method.
The applicability of the method would depend on a physical situation. 
For instance, the number of relevant eigenvalues is small at low temperature, 
while it is large at high temperature due to thermal excitation. 

In the present work, we didn't succeed to solve the inversion of the reduced matrix 
with a shifted term by iterative solvers, which forced us to employ the 
direct method. 
For an efficient implementation of the bSS method, we need to develop
an iterative way to solve the inversion of the reduced matrix, which is a
dense matrix. We would like to address this issue in future studies. 

\section*{Acknowledgement}

This work was supported in part by JSPS Grants-in-Aids for Scientific Research (Kakenhi) 
No. 00586901, 25286097, 60187086, JST CREST, MEXT SPIRE and JICFuS.
The lattice simulations were mainly performed on SX9 at RCNP and CMC at Osaka University.  
This work is also supported by HPCI System Research project (hp130058)
and RICC system at RIKEN.




\end{document}